\documentclass[aps,prl,twocolumn,showpacs]{revtex4}

\pdfoutput=1
\usepackage{hyperref}
\usepackage{bm}

\usepackage{amsmath}
\usepackage{amssymb}
\usepackage{amsthm}
\usepackage{array}
\usepackage{xy}

\usepackage{graphicx}
\usepackage{subfigure}
\usepackage{epstopdf}
\usepackage{grffile}

\usepackage[final]{movie15}
\usepackage{xcolor}

\def\f#1{Fig.~\ref{#1}}

\def\kt{k_{\rm B}T}

\begin{document}

\title{Competing thermodynamic and dynamic factors select molecular assemblies on a gold surface}
\author{Thomas K. Haxton$^{1}$, Hui Zhou$^{2,3}$, Isaac Tamblyn$^{1,4}$, Daejin Eom$^{2,5}$, Zonghai Hu$^{2,6}$, Jeffrey B. Neaton$^{1}$, Tony F. Heinz$^{2,7}$\footnote{tfh3@columbia.edu}, and Stephen Whitelam$^{1}$\footnote{swhitelam@lbl.gov}}

\affiliation{$^1$Molecular Foundry, Lawrence Berkeley National Laboratory, Berkeley, CA 94720, USA; $^2$Department of Physics, Columbia University, New York, NY 10027, USA; $^3$Brion Technologies, Santa Clara, CA 95054, USA; $^4$Department of Physics, Faculty of Science, University of Ontario Institute of Technology, Oshawa, ON L1H 7K4, Canada; $^5$KRISS, Daejeon 305-340, South Korea; $^6$School of Physics, Peking University, Beijing 100871, China; $^7$Department of Electrical Engineering, Columbia University, New York, NY 10027, USA}

\begin{abstract}
Controlling the self-assembly of surface-adsorbed molecules into nanostructures requires understanding physical mechanisms that act across multiple length and time scales.  By combining scanning tunneling microscopy with hierarchical {\em ab initio} and statistical mechanical modeling of 1,4-substituted benzenediamine (BDA) molecules adsorbed on a gold (111) surface, we demonstrate that apparently simple nanostructures are selected by a subtle competition of thermodynamics {\em and} dynamics.  Of the collection of possible BDA nanostructures mechanically stabilized by hydrogen bonding, the interplay of intermolecular forces, surface modulation, and assembly dynamics select at low temperature a particular subset: low free energy oriented linear chains of monomers, and high free energy branched chains.
\end{abstract}

\pacs{64.75.Yz, 81.07.Nb, 68.43.De, 68.43.Bc, 68.43.Hn}

\maketitle

Controlling the self-assembly of surface-adsorbed molecules into nanostructures would enable new classes of functional hybrid interfaces. However, achieving such control requires understanding 
competing physical mechanisms that act across multiple length and time scales
~\cite{elemans2009molecular,Barth2005,Bartels2010,marschall2010random}. In particular, competition between finely balanced intermolecular forces~\cite{blunt2008random,stannard2010entropically} and between intermolecular and molecule-surface forces~\cite{Otero2011, Bartels2010, Gao2010, Kuhnle2009, Barth2007} can lead to rich phase behavior and dynamics~\cite{stannard2011broken,stannard2010entropically,whitelam2012random}. A comprehensive understanding of self-assembly at surfaces therefore requires quantitative connection between molecular forces, nanostructure free energies, and mesoscale assembly dynamics. 

\begin{figure}
\includegraphics[width=0.9\linewidth]{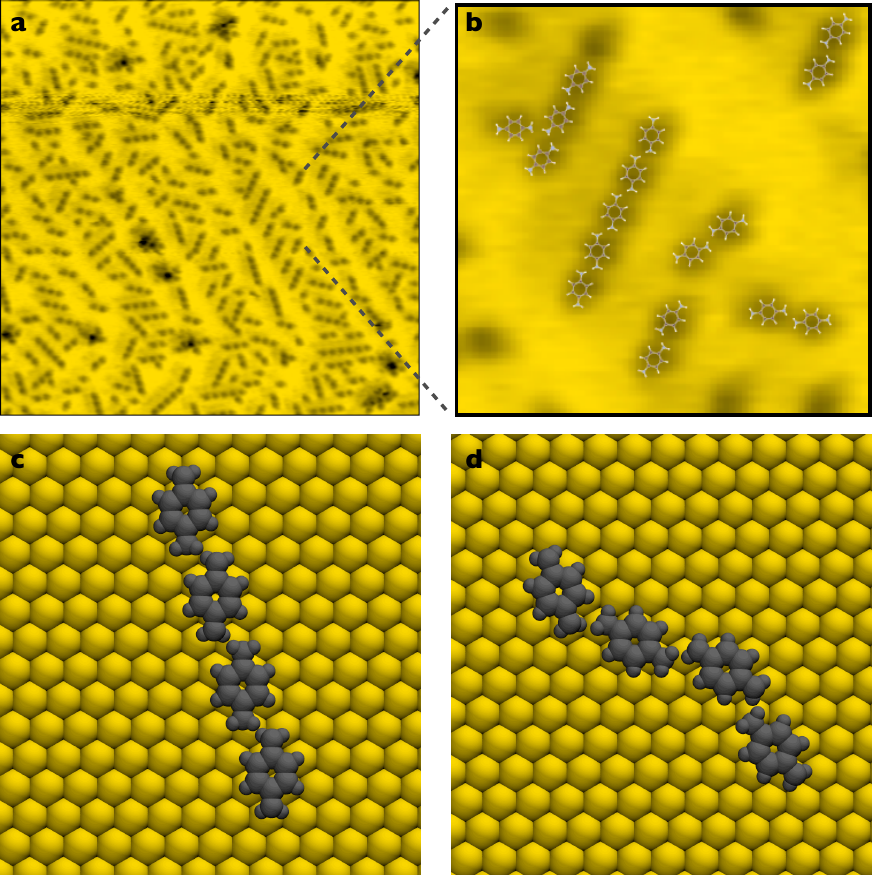}
\caption{{\em Straight chain and branched chain BDA nanostructures are seen in STM images at 5 K}. {\bf a} Experimentally measured STM image of BDA nanostructures on gold (111) show preferential adsorption of nanostructures on high-symmetry regions of the herringbone surface reconstruction.  {\bf b} At increased magnification, energy-minimized DFT configurations overlaid on the experimental images confirm that these nanostructures are stabilized by hydrogen bonds. Of 2481 inspected 3- and 4-molecule nanostructures, 21\% are branched chains and 79\% are straight chains, posing two puzzles: branched chains are much higher in energy than straight chains, and should not be observed in thermal ensembles at low temperature; and {\bf c} straight chains and {\bf d} zigzag chains are equally favored by hydrogen bonding, yet zigzag chains are not seen in experiment. Resolving these puzzles requires identifying the many-body effects that thermodynamically stabilize and dynamically select BDA nanostructures.}
\label{overview}
\end{figure}

The self-assembly of 1,4-substituted benzenediamine (BDA) on a gold (111) surface is an exemplar of the multiscale nature of molecular organization at surfaces, because gold and BDA occupy an important middle ground in the spectrum of intermolecular and substrate interactions: BDA organizes via intermolecular hydrogen bonding~\cite{Colapietro1987}, and gold can impart order to molecular overlayers~\cite{Silly2008, Gao2010, Otero2011,DellAngela2010}. We used high-resolution scanning tunneling microscopy (STM) to image nanostructures self-assembled following gas phase deposition of BDA on the `herringbone' reconstruction of gold (111)~\cite{Wang2007}, and a multiscale simulation approach to determine how nanostructures were selected. Density functional theory (DFT) calculations explain the mechanical stability of local motifs as a consequence of intermolecular hydrogen bonding. A statistical mechanical model -- systematically parameterized using DFT -- reveals how many-body effects resulting from a subtle interplay of molecular forces, surface modulation, and assembly dynamics select at low temperature a subset of possible BDA nanostructures: low free energy oriented linear chains of monomers, and high free energy branched chains.


\f{overview} shows STM images of the experimental system, a collection of immobilized BDA molecules on a gold (111) surface, imaged at 5 K following a temperature quench from 373 K~\cite{PRLsupp}.  These images reveal that molecules 
self-assemble during the quench into a mixture of straight and branched chains, and localize preferentially at high-symmetry regions of the herringbone-reconstructed gold (111) surface. On a coarse scale, the observed preference for high-symmetry regions can be reproduced by assuming the reconstruction imparts an effective oscillatory potential. On a finer scale, the observed chain structures represent locally favorable arrangements of intermolecular hydrogen bonds, illustrated by overlaying a portion of the STM image with energy-minimized DFT configurations of flat BDA molecules (\f{overview}{\bf b}). Further association of nanostructures may be prevented by their sluggish diffusivity, or by energy barriers resulting from the herringbone reconstruction. However, while straight chains (\f{overview}{\bf c}) proliferate, the images are strikingly devoid of zigzag chains (\f{overview}{\bf d}), whose hydrogen bonding is shown by DFT to be energetically equivalent to that of straight chains. Instead, energetically disfavored branched structures are observed. In what follows we show how competitive molecule-surface and molecule-molecule interactions lead to a thermodynamic selection of straight over zigzag chains, as well as a selection of chain orientations, and how sluggish dynamics prevent branched structures from relaxing to the thermodynamically preferred straight chain configuration.

\begin{figure}[t]
\includegraphics[width=0.9\linewidth]{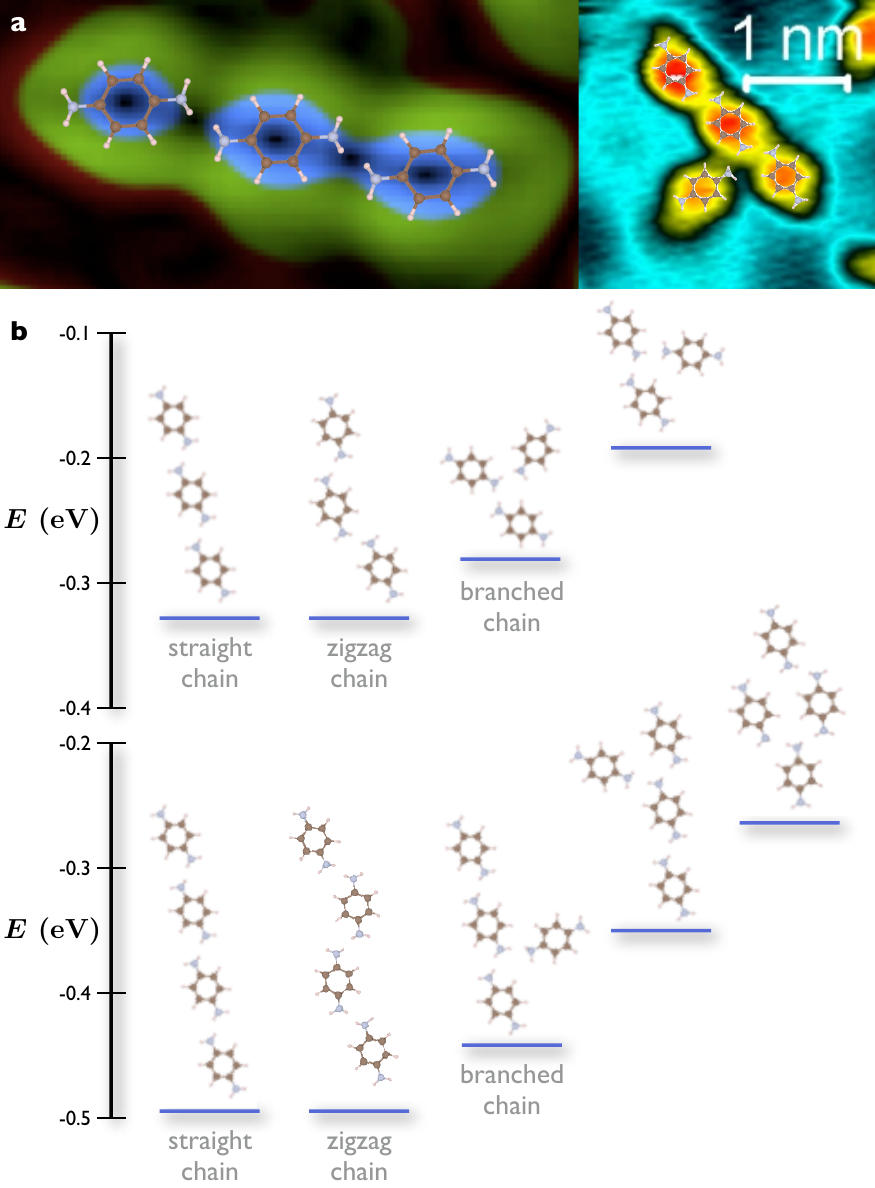}
\caption{{\em Individual BDA nanostructure morphologies are consistent with stabilization by hydrogen bonding, but their relative abundances are not.} {\bf a} Two common experimental motifs overlaid by energy-minimized DFT configurations of flat BDA molecules. The close agreement between experiment and simulation reveals both that chains result from hydrogen bonding between amine groups, and that molecules lie flat on the gold surface. {\bf b} Minimum-energy DFT configurations of 3 and 4 BDA molecules show a clear hierarchy of nanostructure energies. However, STM images show a mixture of high-energy branched chains and low-energy straight chains, while equally low-energy zigzag chains are not seen.}
\label{configurations}
\end{figure}

The morphologies of individual BDA nanostructures result from intermolecular hydrogen bonding. In \f{configurations}{\bf a} we overlay two common experimental motifs with energy-minimized DFT configurations of flat BDA molecules. The 
comparison demonstrates
that motifs result from hydrogen bonding between amine groups, and that molecules lie flat on the gold surface. We found similar agreement for other motifs, including straight chains consisting of as many as 7 molecules whose intermolecular spacing is indistinguishable from that predicted by DFT (see Supplemental Discussion~\cite{PRLsupp}). However, experimental motifs are not observed in the proportion expected from their hierarchy of hydrogen bonding energies (\f{configurations}{\bf b}).  In our sample of 26 STM images, 16\% of three-molecule chains and 28\% of four-molecule chains are branched; based on hydrogen bond energies, we would expect a negligible thermal population of branched structures at 5 K.  Further, all 1969 of the remaining chains are straight, rather than zigzag, despite the fact that straight and zigzag chains have equivalent hydrogen bond energies.


\begin{figure}[t!]
\includegraphics[width=0.5\textwidth]{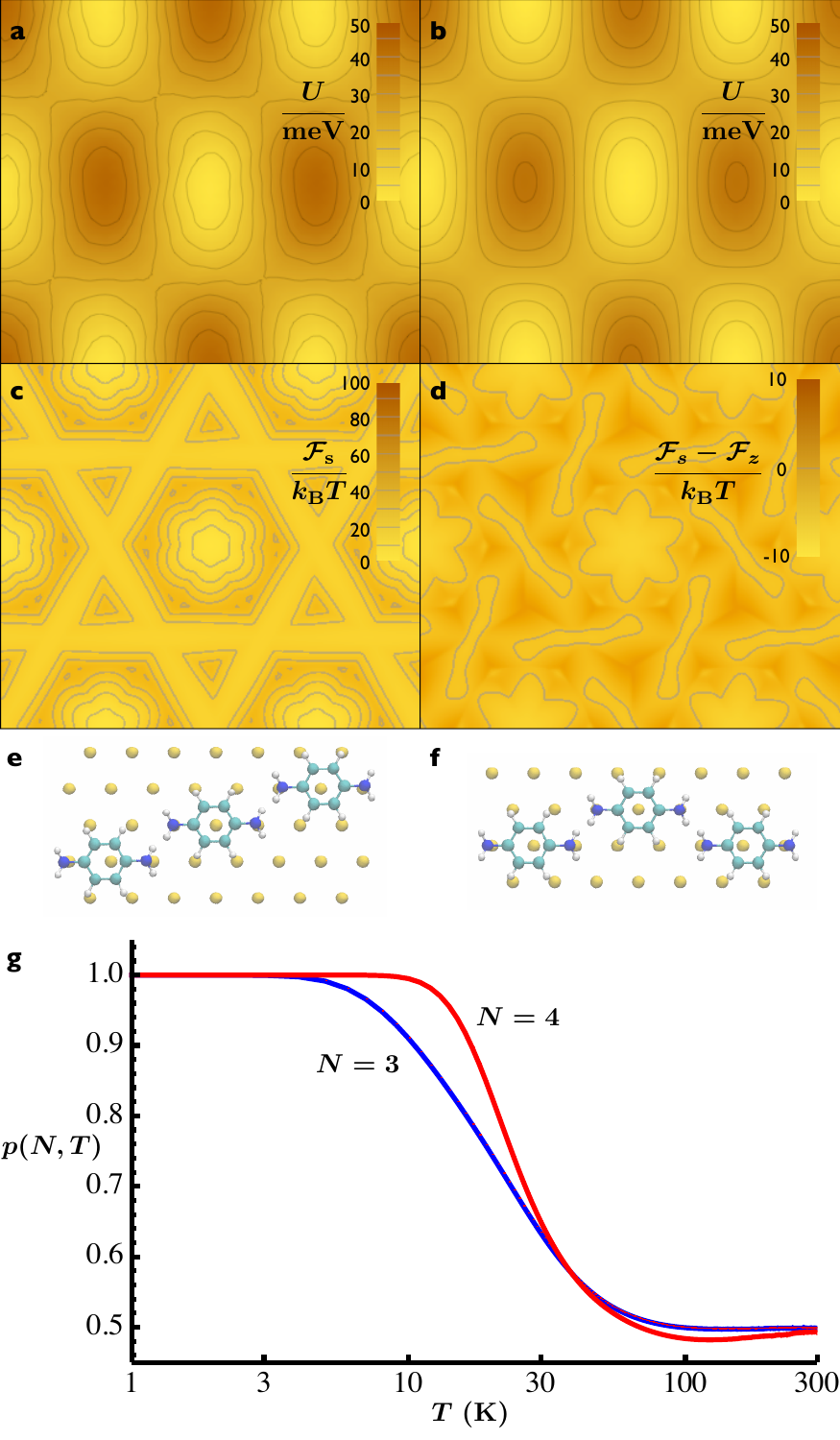}
\caption{{\em Surface modulation thermodynamically favors straight over zigzag chains at low temperature.}  {\bf a} DFT potential energy between a flat BDA molecule and the gold surface. Light colors denote energetically favorable top sites. 
{\bf b} This modulation is reproduced faithfully by our statistical mechanical model, which we used to calculate 
surface-mediated free energies, such as panel {\bf c}, for a three-molecule straight chain at 6.8 K.
Light colors denote low free energies where the middle molecule lies on an energetically favorable top site, as seen in the minimum-energy configurations for {\bf e} straight and {\bf f} zigzag chains.  {\bf d} The free energy difference between straight and zigzag chains illustrates that straight chains are favored near top sites.
{\bf g} Equilibrium preference $p$ for 3-molecule and 4-molecule chains straight chains over their zigzag counterparts, as a function of temperature, calculated using Eq.~(\ref{chaineq}). At low temperature there is a pronounced thermodynamic bias in favor of straight chains.}
\label{straightzig}
\end{figure}

To explain the abundance of branched chains and the absence of zigzag ones we developed a multiscale theoretical approach~\cite{PRLsupp}.  We used DFT to parameterize a three dimensional, statistical mechanical model of BDA on gold. The model includes three pairwise-additive configurational energy contributions: an energy cost for the BDA amine group torsion, an optimized atomistic interaction energy between pairs of BDA molecules (dominated by hydrogen bonding), and an interaction energy between a single BDA molecule and an unreconstructed gold (111) surface. Panels {\bf a} and {\bf b} of \f{straightzig} show one element of the latter, the energy `corrugation' felt by a flat BDA molecule as a function of surface position. This corrugation favors adsorption to top sites. The model calculation reproduces the DFT potential energy surface, at a substantial ($10^8$-fold) reduction of computational cost.


Surface corrugation affects straight and zigzag chains differently. We calculated the effective free energies
\begin{equation}
{\cal F}_{\rm s,zz}(N,T,{\bm r}) = -\kt \ln \int {\rm d} \theta \exp(-\beta U_{\rm s,zz}(N,{\bm r},\theta))
\end{equation}
of dilute straight (s) and zigzag (zz) chains of $N$ molecules, constrained to lie flat on a gold surface at a height that minimizes the BDA-gold energy. Here $U_{\rm s,zz}$ is the interaction energy of a chain with the surface (see Supplemental Methods~\cite{PRLsupp}), $\theta$ is the chain angle, ${\bm r}$ is the position of (one of) the middle molecule(s), and $T=6.8$ K equals the calculated freeze-out temperature for diffusion of single molecules on the surface (see Supplemental Methods~\cite{PRLsupp}; free energies at 5 K have similar forms). \f{straightzig}{\bf c} shows the effective free energies for a straight 3-molecule chain, and \f{straightzig}{\bf d} shows the difference in free energy between straight and zigzag 3-molecule chains, demonstrating how the surface modulation alternately favors straight and zigzag chains. Integrated over position, the net effect of this modulation is 
a temperature-dependent thermodynamic bias in favor of straight chains. We plot in \f{straightzig}{\bf g} the resulting equilibrium preference for straight chains over zigzag ones,
\begin{equation}
\label{chaineq}
p(N,T) = \frac{Z_{\rm s}(N,T)}{Z_{\rm s}(N,T)+Z_{\rm zz}(N,T)},
\end{equation}
where $Z_{\rm s,zz}(N,T) \equiv \int {\rm d} {\bm r} \, \exp(-\beta {\cal F}_{\rm s,zz}(N,T,{\bm r}))$. Although little bias is seen at the high temperatures at which molecules are prepared, there is a strong thermodynamic preference for straight chains at the low temperatures at which STM images are taken. This preference exists because in their lowest energy configurations, BDA molecules in straight chains sit closer to top sites than do molecules in zigzag chains. However, as shown by the minimum-energy configurations of \f{straightzig}{\bf e} and {\bf f} (for $N=3$), this difference, geometrically speaking, is slight: both straight and zigzag chains can adopt configurations with their molecules nearly centered on top sites (see Supplemental Discussion~\cite{PRLsupp}). Thus, entropy overwhelms the energetic preference for straight chains except at low temperature, where the surface modulation is large enough compared to $k_{\rm B} T$ to induce a bias in favor of straight chains.  Surface corrugation also imparts a preference for absolute chain angle (see Sec. 3 and Fig. 3 of the Supplemental Discussion~\cite{PRLsupp}).

\begin{figure}
\includegraphics[width=0.5\textwidth]{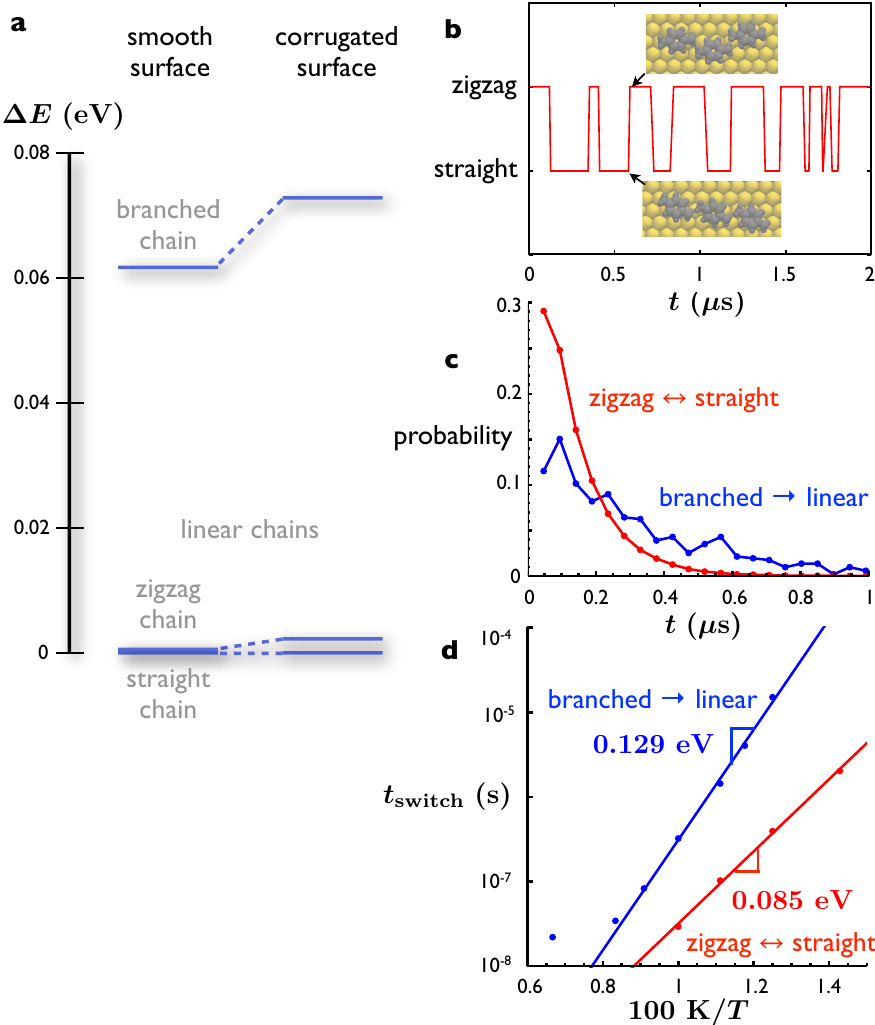}
\caption{{\em The dynamics of linear chains is much faster than that of branched chains, suggesting that the latter are seen experimentally because they are kinetically trapped.} {\bf a} Relative free energies of branched, zigzag, and straight chains of length 3 on a smooth surface (left) and a gold (111) surface (right). High-energy branched chains are seen in STM images, while low-energy zigzag chains are not. {\bf b} Section of a typical dynamic trajectory in equilibrium at $T=100$ K, showing multiple switching events between zigzag and straight chains. Inset: snapshots of configurations immediately before and after a switching event. {\bf c} Distribution of switching times from many simulations, combined with data showing the slower conversion of branched chains to linear ones. {\bf d} Average first switching time from branched chain to linear chain (blue curve), and average switching time between zigzag and straight chains (red curve) as functions of inverse temperature.  Lines represent best fits to Arrhenius rate laws with characteristic energy barriers. The sluggishness of branched chain relaxation at low temperature suggests that they are a kinetically trapped species.}
\label{dynamics}
\end{figure}

Although surface modulation explains the preference for straight over zigzag chains, surface modulations acts to further {\em disfavor} branched chains with respect to the linear chain types (Fig.~\ref{dynamics}{\bf a}). Why, then, are so many branched chains seen in STM images? The answer 
stems from the distinctly different dynamics of branched and linear nanostructures. To efficiently approximate the dynamics of our statistical mechanical model, we carried out dynamical simulations using the virtual-move Monte Carlo algorithm~\cite{Whitelam2007}, which moves individual molecules and clusters self-consistently according to gradients of potential energy acting on them.  Although the algorithm was developed to model overdamped motion in solution, here we parameterized it to approximate stochastic but underdamped motion induced by thermal vibrations of the gold substrate~\cite{PRLsupp}.  As shown in Fig.~\ref{dynamics}{\bf b}, dynamical simulations at 100 K, above the chain selection temperature, display frequent switching of linear chains between straight and zigzag modes. The associated distribution of switching times is shown in Fig.~\ref{dynamics}{\bf c}. Branched nanostructures, by contrast, convert to linear nanostructures much more slowly (Fig.~\ref{dynamics}{\bf c}). 

The relative conversion rates revealed by direct dynamical simulation (Fig.~\ref{dynamics}{\bf d}) indicates a greater propensity for branched chains to become kinetically trapped.  Branched chains must rupture hydrogen bonds to convert to linear chains (see Supplemental Movie 1~\cite{PRLsupp}), explaining why the 0.129 eV activation energy of Fig.~\ref{dynamics}{\bf d} approaches the 0.170 eV hydrogen bond strength. By contrast, zigzag and straight chains can interconvert without breaking contact between amine groups (see Supplemental Movie 2~\cite{PRLsupp}), resulting in an energy barrier of only 0.085 eV. 
At low temperatures,
we expect this difference in interconversion rates to be even greater: branched chains must still break hydrogen bonds in order to relax, but linear chains may interconvert along more tortuous but energetically cheaper paths than those seen in dynamical simulation at about 100 K.  We used numerical minimization~\cite{Weinan2007} to identify a minimum-energy barrier for zigzag-to-straight chain relaxation of 0.041 eV, suggesting an even greater disparity between relaxation rates of branched and zigzag structures at low temperatures.  As the system is cooled, therefore, we expect branched nanostructures to become kinetically trapped before their linear counterparts. We expect all species to be frozen at the observation temperature of 5K.

Our results are qualitatively consistent with the following physical picture: 1) zigzag chains relax to the straight chain structure preferred at low temperature by the gold substrate~\footnote{The linear chain selection mechanism may be relevant for other assemblies that show a preference for straight- over zigzag chains~\cite{Kwon2009}.}, while 2) branched chains are kinetically trapped, and cannot convert to the energetically preferred linear chain types.  As described in the Supplemental Discussion~\cite{PRLsupp}, a master equation approach incorporating extrapolated conversion rates and calculated free energies yielded dynamics qualitatively consistent with this picture: branched chains remain kinetically trapped after a temperature quench, while zigzag chains relax preferentially to straight chains. However, our master equation approach predicted an appreciable population of zigzag chains, in contrast to their experimentally observed absence.  A detailed sensitivity analysis~\cite{PRLsupp} illustrates how this discrepancy may result from small numerical uncertainties in calculated and fitted energies.
Notably, the uncertainties inherent in 
using approximate DFT functionals,
although small numerically, are significant for the current problem.  For example, the free energy difference between straight and zigzag chains is proportional to the absolute magnitude of the gold surface corrugation, which depends on the adsorption height.  When the adsorption height is varied by 0.5 \AA~from its energy-minimized value to reflect known inaccuracies in adsorption heights from state-of-the-art, van-der-Waals-corrected DFT functionals~\cite{Li2012} (see Supplemental Discussion~\cite{PRLsupp}), the free energy difference varies by 0.004 eV.  Although this variation is small, 
it is sufficient to significantly change the population of zigzag chains predicted by the master equation approach.
This sensitivity to numerical uncertainties illustrates how quantitative description of the type of nanostructure selection seen in our experiments remains a significant challenge for theoretical methods.


The apparent simplicity of BDA nanostructures seen in the low-coverage limit on a gold (111) surface belies the complex competition of forces and dynamics that selects them.  We have combined density functional theory and statistical mechanics to show that while intermolecular hydrogen bonding determines the morphologies of possible nanostructures, a combination of surface-mediated thermodynamics and dynamics selects which structures self-assemble at low temperature in experiments.  Straight and zigzag chains are equally stable at high temperature, but straight chains are stabilized by the gold surface at low temperature. Branched nanostructures are less stable then both of these, but their sluggish dynamics at low temperature prevents their relaxation to the stable linear forms.  Our results show how kinetic factors can complement thermodynamic ones so as to yield greater control over the relative abundances of molecular nanostructures; we predict that such kinetic control could be achieved for the present system using a temperature-controlled experimental set-up.  However, fully quantitative prediction of nanostructure populations requires a precision that exceeds that of current theoretical methods. The 
further development of this
and other multiscale methods~\cite{Mannsfeld2006,noid2008multiscale,noid2008multiscale2,krishna2009multiscale4,noid2009multiscale,chan2005simulations,mccabe2004multiscale,grunwald2012transferable} is urgently required in order to predict the assembly behavior of molecules on surfaces generally~\cite{Otero2011, Bartels2010, Kuhnle2009, elemans2009molecular, Barth2007, Barth2005}, whose complex mesoscopic behaviors can result from a subtle competition of microscopic forces and dynamics.

\begin{acknowledgments} 

This research was done as part of a User project at the Molecular Foundry, supported by the Office of Science, Office of Basic Energy Sciences of the U.S. Department of Energy under contract DE-AC02-05CH11231. Work at Columbia was supported by DOE contract DE-FG02-07ER15842.
Simulations and DFT calculations used resources of the National Energy Research Scientific Computing Center, which is supported by the Office of Science of the U.S. Department of Energy under Contract No. DE-AC02-05CH11231. The authors thank Prof. George Flynn for helpful discussions.

\end{acknowledgments}



\end{document}